\documentclass[sigconf]{aamas} 

\renewcommand\footnotetextcopyrightpermission[1]{} 
\usepackage{balance} 
\usepackage{booktabs}  
\usepackage{caption}   
\usepackage{siunitx}   
\usepackage{amsmath,amsfonts}
\usepackage{bm}
\usepackage{nicefrac,xfrac}
\usepackage[linesnumbered,ruled,vlined]{algorithm2e}
\usepackage{graphicx}
\usepackage{hhline}
\usepackage{textcomp}
\usepackage{xcolor}
\usepackage{tabularx}
\usepackage{multirow}

\usepackage[capitalize,noabbrev,nameinlink]{cleveref}
\crefformat{equation}{(#2#1#3)}
\Crefname{algocfline}{Algorithm}{Algorithms}
\Crefname{algocf}{line}{lines}
\crefname{table}{Table}{Tables}
\Crefname{table}{Table}{Tables}
\Crefname{question}{Question}{Questions}
\Crefname{assumption}{Assumption}{Assumptions}
\crefrangeformat{assumption}{Assumptions~#3#1#4-#5#2#6}

\DeclareMathOperator*{\argmin}{arg\,min}

\setcopyright{ifaamas}
\acmConference[AAMAS '25]{Proc.\@ of the 24th International Conference
on Autonomous Agents and Multiagent Systems (AAMAS 2025)}{May 19 -- 23, 2025}
{Detroit, Michigan, USA}{A.~El~Fallah~Seghrouchni, Y.~Vorobeychik, S.~Das, A.~Nowe (eds.)}
\copyrightyear{2025}
\acmYear{2025}
\acmDOI{}
\acmPrice{}
\acmISBN{}

\acmSubmissionID{1101}

\title[]{Improving DeFi Mechanisms with Dynamic Games and Optimal Control: A Case Study in Stablecoins}

\author{Nicholas Strohmeyer}
\affiliation{
  \institution{The University of Texas at Austin}
  \city{Austin}
  \country{United States}}
\email{nstrohmeyer@utexas.edu}

\author{Sriram Vishwanath}
\affiliation{
  \institution{The University of Texas at Austin}
  \city{Austin}
  \country{United States}}
\email{sriram@utexas.edu}

\author{David Fridovich-Keil}
\affiliation{
  \institution{The University of Texas at Austin}
  \city{Austin}
  \country{United States}}
\email{dfk@utexas.edu}

\begin{abstract}
Stablecoins are a class of cryptocurrencies which aim at providing consistency and predictability, typically by pegging the token's value to that of a real world asset. Designing resilient decentralized stablecoins is a challenge, and prominent stablecoins today either (i) give up on decentralization, or (ii) rely on user-owned cryptocurrencies as collateral, exposing the token to exogenous price fluctuations. In this latter category, it is increasingly common to employ algorithmic mechanisms to automate risk management, helping maintain the peg. One example of this is Reflexer's RAI, which adapts its system-internal exchange rate (redemption price) to secondary market conditions according to a proportional control law. In this paper, we take this idea of active management a step further, and introduce a new kind of control scheme based on a Stackelberg game model between the token protocol and its users.  By doing so, we show that (i) we can mitigate adverse depeg events that inevitably arise in a fixed-redemption scheme such as MakerDao's DAI and (ii) generally outperform a simpler, adaptive-redemption scheme such as RAI in the task of targeting a desired market price. We demonstrate these results through extensive simulations over a range of market conditions.
\end{abstract}


\newcommand{\BibTeX}{\rm B\kern-.05em{\sc i\kern-.025em b}\kern-.08em\TeX}

\newcommand{\pstb}{{p}^{\textnormal{Stb}}_{t}}
\newcommand{\ps}{\widehat{p}^{\textnormal{Stb}}_{t}}
\newcommand{\pC}{\widehat{p}^{C}_{t}}

\newcommand{\peg}{p^{\textnormal{peg}}}
\newcommand{\JP}{J^{\textnormal{Ptcl}}}
\newcommand{\JS}{J^{\textnormal{Spec}}}
\newcommand{\US}{U^{\mathrm{Spec}}}
\newcommand{\err}{e_t}
\newcommand{\epeg}{e_t^{\textnormal{peg}}}
\newcommand{\blam}{\bm{\lambda}}
\newcommand{\bmu}{\bm{\mu}}
\newcommand{\beps}{\bm{\epsilon}}

\newcommand{\timeT}{t \in \{ 0, 1, \dots , T  \} }
\newcommand{\tol}{\mu_{\textnormal{tol}}}
\newcommand{\zlast}{\bm{z^{*}_{t-1}}}
\newcommand{\zstar}{\bm{z^{*}}}

\SetCommentSty{mycommfont}

\begin{document}


\pagestyle{fancy}
\fancyhead{}


\maketitle 


\section{Introduction}

Decentralized Finance (DeFi) utilizes blockchain technology to provide financial services through secure, peer-to-peer transactions, eliminating the need for intermediaries \cite{werner2022sok}. Although DeFi has significant potential, many cryptocurrencies experience high volatility, reducing their effectiveness as a reliable medium of exchange. Stablecoins, a subset of digital currencies, are designed to counter this volatility by pegging their value to real-world assets like the US dollar. However, maintaining a stable peg while preserving decentralization presents a challenge, typically addressed through two flawed approaches. The first method involves backing the stablecoin with exogenous, on-chain assets, which are often volatile cryptocurrencies that introduce additional risk. The second method employs algorithmic mechanisms to maintain the peg via automated monetary policy, but these systems can be fragile and have failed during crises \cite{clements2021built}.


A key component of stablecoin systems is the process of token creation (``minting'') and redemption (``burning''). This usually involves a transaction between the protocol and participants, where a basket of collateral assets is exchanged for new tokens, or vice versa. When the rate of minting and burning aligns with market supply and demand, the peg is maintained. However, if there is a significant imbalance in this exchange, the stablecoin risks ``depegging.''

To address such imbalances, the protocol may adjust the endogenous rate at which participants can exchange stablecoin tokens for the underlying collateral assets—referred to as the \emph{redemption price}. This alters the incentive structure for participants. For example, increasing the redemption price raises the future expected cost of recovering collateral, encouraging participants to burn, while lowering the price has the opposite effect. By modulating the redemption price, the protocol can indirectly influence token supply dynamics.

Several past stablecoins have implemented variable redemption price mechanisms \cite{rai, lauko2021liquity, gyro, adams2022runs}. A particularly relevant example is Reflexer's RAI \cite{rai}, which uses a feedback control law to continuously adjust based on market conditions. Reflexer's controller updates the redemption price at a rate proportional to its current deviation from the market price, allowing RAI to maintain relative stability while staying fully decentralized. This contrasts with its predecessor, DAI, which uses a fixed redemption price and has increasingly had to rely on fiat-backed assets in recent years \cite{psm}. However, RAI does not target a fixed peg value; instead, its goal is to align the redemption and market price movements. Despite this, RAI's success highlights the potential of feedback control mechanisms in stablecoin systems.   

\textbf{Our main contribution} in this paper is to introduce a novel feedback control scheme for dynamically updating the redemption price of a decentralized stablecoin system. Like Reflexer's RAI, we incorporate feedback signals from the market to inform future decisions. Unlike Reflexer, we leverage domain-specific insights to model the supply dynamics of decentralized stablecoins, enabling us to use techniques from \emph{optimal} control. Specifically, we capture the effect of protocol redemption price decisions on the behavior of system participants as a dynamic {\em Stackelberg game} \cite{li2017review, von2010market} and formulate this as a mathematical program with equilibrium constraints \cite{luo1996mathematical}. Lastly, we demonstrate how to efficiently find first-order equilibrium solutions to the resulting optimization problem.

To assess our design, we implement it using a {\em receding horizon} approach, re-solving for the optimal sequence of redemption rates over a finite future window at each time step, and applying the first rate from that sequence. We compare this approach against two existing methods: a fixed-rate scheme based on DAI and a proportional control scheme based on RAI. Our results show that the Stackelberg controller reduces the volatility of the stablecoin's market price in simulations relative to an arbitrary peg value and enables faster recovery from market shocks when compared to the two baselines.

\section{Background}



The breakthrough innovations of Bitcoin in 2009 and Ethereum in 2015 led to the emergence of DeFi. The former ushered in a new way of conducting peer-to-peer transactions using blockchain technology while the latter introduced smart contracts, allowing for specific sets of rules to be executed automatically on-chain. DeFi has since grown into a complex, integrated system of applications supporting many traditional functions of institutional finance \cite{jensen2021introduction, schueffel2021defi}.
In 2014, Tether introduced the first stablecoin (USDT) which uses the classic model of targeting a 1:1 peg to the US dollar. More broadly, a stablecoin can be pegged to a variety of real world assets, such as other fiat currencies, commodities, precious metals, real estate, etc. \cite{mita2019stablecoin}. Stablecoins play a critical role in the emerging crypto-economy since they serve as safe-haven assets \cite{grobys2021stability, ante2021influence}. 

\subsection{Stablecoin Design Paradigms} 
While most stablecoins have the same (or similar) end goals, their designs can vary drastically. To understand the many designs that exist today, a useful, first distinction is to separate \emph{custodial} from \emph{non-custodial} stablecoins \cite{klages2020stablecoins, d2021stablecoins}.

\noindent\textbf{Custodial Stablecoins} rely on a central entity to maintain reserves of real-world assets from which the on-chain token derives its stability. This is generally effective, but a key drawback is that decentralization is sacrificed in order to maintain the peg. Also, backing a stablecoin with a reliable fiat currency is still no guarantee of stability and there has been history of these tokens temporarily losing their peg \cite{svb, moodyb, moodya}. Despite this, custodial tokens dominate the stablecoin market where today Tether's USDT and Circle's USDC make up roughly 86\% of total market cap \cite{coinmarket}.


\noindent\textbf{Non-custodial stablecoins}, on the other hand, seek to maintain a peg while preserving system decentralization. These stablecoins can be further broken down into two sub-categories corresponding to the strategies used for supporting the peg.

\emph{Crypto-backed stablecoins} maintain decentralization by using cryptocurrencies as system collateral rather than resorting to the safer choice of fiat currencies. However this increases risk exposure, and the system must rely on several additional design features to help mitigate this. 

\emph{Algorithmic stablecoins} rely \emph{solely} on incentive-based mechanisms and endogenous system value (``seigniorage shares") to maintain their peg. They have become somewhat notorious due to several past depegs and failures (TerraUSD \cite{liu2023anatomy}, NuBits\cite{nubits} and Iron\cite{adams2022runs}, among others). Such events have called into question the viability of total reliance on algorithmic mechanisms \cite{clements2021built} and have motivated the growing popularity of so-called \emph{hybrid} models \cite{kazemian2022frax, rai}, which use a combination of both crypto-backing and algorithmic operations. 


\subsection{Crypto-backed Stablecoins (DAI) } 
\label{dai_section}

MakerDAO’s DAI \cite{team2017dai} is a leading example of a crypto-backed stablecoin model and serves as a useful case to highlight key design principles, risks, and mitigation strategies. In its initial version, DAI was backed solely by Ethereum. To manage Ethereum’s price volatility, MakerDAO implemented an {\em over-collateralization} system, meaning that the total market value of the collateral assets exceeds the value of the tokens in circulation.

DAI’s minting process functions as a lending agreement between the protocol and participants. The main incentive for participants, which we refer to as “speculators,” is to make leveraged bets on the long-term value of Ethereum. They do this by opening a collateralized debt position (CDP) with MakerDAO, which requires them to deposit initial collateral worth $\beta N$ (where $\beta > 1$ is the {\em minimum collateral ratio} set by MakerDAO) to mint $N$ tokens. This mechanism distributes system collateral ownership across multiple speculators and ensures over-collateralization is maintained when positions are created.



The speculator's assets are locked in the CDP (also referred to as a ``vault") until they are either redeemed or liquidated. The latter occurs if ever the speculator's vault falls below the minimum required ratio $\beta$, in which case the vault is auctioned off (typically at a loss). Vault redemption is the reverse process of creation in which a speculator recovers their collateral by returning the borrowed DAI (usually plus some interest e.g. "stability fee" in MakerDao). Specifically, in a single-collateral system, when the speculator returns some fraction of the $N$ borrowed tokens, they recover the proportionate quantity of the collateral asset invested.

\subsection{The Role of the Redemption Price}
 Consider the decision faced by the speculator. Assuming they sell new tokens created from the process above to make leveraged bets on ETH, then, when they decide to redeem, they will need to buy those tokens back at the market price $\pstb$. Supposing $\pstb > \$1$, this could result in a loss if they borrowed tokens at a redemption price of \$1 (as is the case with DAI). However, the opposite is true when $\ps < \$1$, where the difference is in their favor.
 





\subsection{Deleveraging Risk and Dynamics}
\label{deleveraging}
Collateral devaluation due to market fluctuations is a primary source of risk in crypto-backed stablecoins.
Prior works \cite{klages2021stability, klages2022while} have highlighted a particular phenomenon, known as a ``deleveraging spiral," which can occur when a large number of vault-owners face immediate risk of liquidation. This triggers a spike in market demand as these owners rush to buy back the tokens needed to settle their positions. Interestingly, \cite{klages2021stability} predicted this pattern prior to these exact dynamics playing out in the MakerDao system during the infamous ``Black Thursday" crisis.
The authors also pointed out that reducing the cost of redemption during extreme deleveraging could alleviate buying pressure and that such a technique could be used as a mitigation strategy.



\section{Preliminaries and Related Work}

This section provides a formal definition of proportional-integral-derivative (PID) control and how it is used specifically in Reflexer's autonomous redemption price setting mechanism. This discussion motivates the use of \emph{optimal control} as an alternative. We also highlight several key related works which have studied similar problems in the context of DeFi protocols, token based economies and leader-follower economic models. 

\subsection{PID Control}
\label{pid section}
Proportional-integral-derivative (PID) control is a feedback control technique which determines the input to a dynamical system in proportion to an ``error'' signal $e$, its integral over time, and its time derivative. 
In discrete-time, this takes the following form:
\begin{equation}
    \label{pid}
    \delta\alpha_t = K_p \err + K_{I}\sum_{\tau=t-T}^{t} e_{\tau} + K_d \biggl(\err - e_{t-1}\biggr )\,
\end{equation}

Reflexer's controller uses only the left-most (proportional) term, dropping the integral and derivative terms. This can be described by $\delta\alpha=K_{p}\err$, where the control input $\delta\alpha_t$ is the \emph{rate of change} to the redemption price $\alpha_t, K_p > 0$ is the constant gain factor set by the protocol, and $e_t = \alpha_t - \pstb $ is the difference between the redemption price $\alpha_t$ and the market price of the stablecoin. 

The goal of this design is to drive the difference between $\pstb$ and $\alpha_t$ smoothly to 0. For instance, when $\pstb > \alpha_t \rightarrow \err < 0$, the derived redemption rate $\delta\alpha_t$ is negative, and $\alpha_t$ decreases proportionally to the magnitude of $\err$. As $\alpha_t$ decreases, system debt becomes cheaper and the incentive for speculators to mint tokens increases. Meanwhile token holders in the secondary market have an incentive to sell as they expect the near-future market value to decrease. These forces combine to push $\pstb$ back towards $\alpha_t$. In the limit $t \to \infty$, both $e_t, \delta\alpha_t \rightarrow 0$ and a temporary equilibrium is met where $\pstb=\alpha_t$. The opposite pattern occurs when $\pstb < \alpha_t$.


In addition to Reflexer, there are other examples of proportional (P) and proportional-integral (PI) control across DeFi \cite{feedback} such as in the Mars \cite{mars} and Euler \cite{euler} lending protocols. 
While they are straightforward to implement and computationally efficient, PID controllers are not built around any notion of optimality, leaving an opening for improvement.

\subsection{Optimal Control} 
In this work, we design a redemption price controller in terms of an optimization problem. This is inspired by past works \cite{akcin2022control, chitra2022defi, sanchez2019truthful, bertucci2024agents} which have approached related DeFi problem settings from the perspective of \emph{optimal control.} 
Optimal control enables designers to encode their goals, such as minimizing fluctuations around the peg, into tailored cost functions  while adding system dynamics and key features, such as specified bounds, as constraints. 

The resulting controller is not just reactive (as is the case for PID) but also \emph{predictive}.
In particular, it can make use of arbitrarily complex forecasting models and solve a corresponding optimal control problem over a user-specified time interval extending into the future. 
Meanwhile, by re-solving the forward-looking problem at each time $t$, the resulting controller is still always responsive to new market observations.


To illustrate, consider that token demand is trending up and is expected to continue increasing. In this case, the protocol can decide to reduce the \emph{current} redemption price to preemptively keep pace with the anticipated growth. On the other hand, when the realized demand trajectory inevitably differs (ideally very little) from the forecast, the solution to the optimization problem at the next step corrects for the difference. This means feedback and short-term prediction are integrated into every control decision, a powerful advantage of this approach over PID control.

The technique just described is often referred to as receding horizon or model predictive control \cite{allgower2004nonlinear,garcia1989model}, and it is widely used, appearing in both engineering \cite{schwenzer2021review,qin2003survey} and financial applications \cite{trimborn2017portfolio,herzog2006model,herzog2007stochastic}. Despite this, it is relatively unexplored in DeFi, with the exception of a few notable works which have demonstrated its utility in designing algorithmic policy \cite{chitra2022defi,sanchez2019truthful} and setting interest rates in lending/borrowing applications \cite{bertucci2024agents}.

\subsection{Stackelberg Equilibrium Model}
Implementing the method from the preceding section requires an effective model of decentralized stablecoin dynamics.
Recent work \cite{akcin2022control} shows that the evolution of a token-based economy can be accurately captured by a dynamical system driven by the decisions of a central controller and its primary participants. 
Analogously, the supply dynamics of a CDP-backed stablecoin are driven by the primary market interaction between the protocol and speculators. 
Inspired by this, we choose to model the interaction as a dynamic Stackelberg game \cite{von2010market} where the protocol acts as a \emph{leader}, which sets the redemption price at each time $t$, and the speculators, modeled as a single \emph{follower}, who burn and mint based on this price in an effort to maximize their expected wealth.

Game-theoretic, leader-follower relationships are naturally structured as bilevel optimization problems \cite{dempe2020bilevel, bard2013practical, akcin2022control} in which a subset of decision variables of an \emph{upper-level} are constrained to lie within the solution set of a \emph{lower-level} problem. Bilevel optimization appears frequently in economic settings as it can capture this notion of agents responding to common, central incentives. Some applications include optimal pricing \cite{brotcorne2001bilevel, labbe2016bilevel}, efficient network design \cite{marcotte1986network, josefsson2007sensitivity}, and energy market models \cite{zugno2013bilevel,yu2016supply} among many others \cite{kalashnikov2015bilevel}. 


\section{Methods} \label{methods}

We now formalize the dynamic Stackelberg game model of a decentralized crypto-backed stablecoin.
At each step $t$, the key quantities---or \emph{state} variables---update according a discrete-time dynamical system driven by net burns/mints $\Delta_t$ and the \emph{exogenous} market price of collateral $p^{C}_{t}$. The speculator's decision $\Delta_t$ is affected by the current redemption price $\alpha_t$; the protocol anticipates this, accounting for the speculator's best response in its own decision for $\alpha_t$. This dynamic interaction is organized into a single bilevel optimization problem which is solved over a finite time horizon of $T \in \mathbb{N}$ future steps. 

\renewcommand{\arraystretch}{1.3} 
\begin{table}[tb]
\centering
\begin{tabularx}{\columnwidth}{|c||X|}
    \hline 
    \textbf{Symbol}& \multicolumn{1}{c|}{\textbf{Definition}} \\
    \toprule \hline
    \(S_t\)                         & Total stablecoin supply at time $t$  \\  \hline
    \(\Delta_t\)                    & Change in supply; speculator's burn/mint decision  \\  \hline
    \(\widehat{D}_t\)               & Market demand forecast $T$ steps ahead of current $t$ \\   \hline
    \(\alpha_t\)                    & System redemption price, protocol's state variable  \\  \hline
    \(\delta\alpha_t\)              & One-step change in \(\alpha_t\), protocol's decision variable \\ \hline
    \(C_t\)                         & Total quantity of collateral assets in CDP's at $t$   \\   \hline
    \(p^{C}_t , \widehat{p}^{C}_t\) & Collateral asset market price / forecast   \\ \hline
    \( \pstb , \ps  \)              & Stablecoin market price / forecast    \\  \hline
    \(r_t \)                        & Collateral one-step returns $\bigl( \nicefrac{p^{C}_{t+1}}{p^{C}_{t}} \bigr) $ \\ \hline
    \(\Gamma_t \)                   & System Collateralization Ratio at $t$: $\nicefrac{ \left(C_t  p^{C}_{t}\right)}{\left(\alpha_t S_t\right)} $  \\ \hline
\end{tabularx}
\label{tab:vars}
\end{table}

\subsection{Protocol Objective}

The protocol's main objective is to maintain the target $\peg$. This is translated into minimizing a sum of squared errors (\(\epeg\)) between predicted market stablecoin prices ($\ps$) and the target price ($\peg$) at each step $\timeT$, where \(\epeg = \left[\ps(\widehat{D}_t,S_t) - \peg\right] \). Meanwhile, the protocol's \emph{control} variable is the step-wise \emph{change to redemption price} $\delta\alpha$. Motivated by Reflexer \cite{rai}, we model the speculator's reaction to $\delta\alpha$ as a proportional response and thus introduce a bi-linear term $\delta\alpha \epeg $ to the protocol's cost function. Generally desiring smooth updates to the redemption price, we add a term which penalizes large values of $\delta\alpha$ by weight $\omega_p$. In full, the protocol's cost is given by the following:
\begin{equation}
    \label{eqn:protocol}
    \JP = \sum_{t=0}^{T} \left[ (\epeg)^2 + \omega_p \cdot \delta\alpha_t^2 + \delta\alpha_t \cdot \epeg \right].
\end{equation}

\emph{Remark}: In our implementation of \cref{eqn:protocol}, we adapt $\omega_p$ to relax the penalty whenever $\epeg$ falls outside a tolerance threshold. For example, we chose that if $|\epeg| > .01$, then $\omega_p = 1$, otherwise $\omega_p = \nicefrac{1}{|\epeg|}$.

\subsection{Speculator Objective}
\label{specobjective}
The speculator's control variable is the change in token supply ($\Delta_t$) at time $t$ where the sign of $\Delta_t$ indicates whether the action was a net mints or net burn. 
We model the decision as a utility maximization problem. In  \cite{klages2021stability}, the authors model the speculator's \emph{expected} long-term extractable wealth after a single decision $ \Delta_t $ as follows:
\begin{equation}
    \label{wealth}
    W_{t+1} = r_t\left( C_tp^{C}_{t} + \pstb \Delta_t \right) - \alpha_t(S_t +\Delta_t)\,.
\end{equation}
Inspection of \cref{wealth} reveals that $W_{t+1}$ is the value of the speculator's collateral assets at $t$ minus their liabilities (the token supply at the current redemption level $\alpha_t$). 

Next, we assume the speculator acts as if maximizing $W_{t+1}$ by greedily maximizing the sum of discounted marginal gains at each step. The marginal gain at $t$ can be derived from \cref{wealth} by dropping the terms which do not depend on $\Delta_t$. By replacing $r_t, \pstb$ with forecasts $\widehat{r}_t, \ps$ and adding the discount factor $\gamma$, we obtain the following:

\begin{equation}
    \label{eqn:utility}
    \US = \sum_{t=0}^{T-1} \gamma^{t} \left( \widehat{r_t}\ps - \alpha_t \right) \Delta_t\,.
\end{equation}
Lastly, in a system with an adaptive redemption price, the speculators will naturally make arbitraging trades tending to align \(\ps\) with $\alpha_t$. We encode this notion as a quadratic ``penalty" term, which the speculator tries to minimize. The overall utility is thus given as follows:

\begin{equation}
    \label{eqn:utility}
    \US = \sum_{t=0}^{T-1} \gamma^{t} \left( \widehat{r_t}\ps - \alpha_t \right) \Delta_t + w_{S}\left[\alpha(t) - p(\Delta_t)\right]^2  \
\end{equation}
where $w_{S}$ is a configurable weight parameter, which we set to 1. In summary, the first term represents a greedy maximization of long term wealth and depends on \emph{exogenous} collateral prices, while the second term represents presence of arbitrageurs capitalizing on differences between the endogenous redemption price and the token’s market price.



\subsection{System Dynamics}
State variables evolve over a single transition, driven by agent decisions 
\begin{subequations}
    \begin{align}
    \alpha_{t+1} =& \alpha_{t} + \delta\alpha_{t} \label{rates} \\
     S_{t+1} =& S_t + \Delta_t \label{supply} \\
     C_{t+1} =& C_t + \frac{\pstb}{p^{C}_{t}}\Delta_t \label{collateral}
     \end{align}
\end{subequations}

The ratio $\nicefrac{\ps}{p^{C}_{t}}$ is the instantaneous conversion factor between the token and the collateral asset at time $t$. Equation \cref{collateral} therefore implies that, whenever the speculator burns or mints stablecoins, they are instantaneously converting to or from the collateral asset.

Constraint \cref{rates,supply,collateral} are central to the model predictive approach, as they allow for a \emph{trajectory of actions} to be jointly optimized over a horizon of $T$ time steps into the future. Since updates depend on future market observations ($\pstb, p^{C}_{t}$), which are unknown at present (time $t=0$), we replace the actual values with forecasts $\ps, \pC$.

\subsection{Minimum Vault Constraint}\label{vault} 
The vault collateralization ratio is given by $\Gamma_t = \nicefrac{C_t p^{C}_t}{\alpha_{t}S_t}$. The speculator must always ensure $\Gamma_t \geq \beta_t$ to avoid liquidation. As is the case with DAI \cref{dai_section}, we set a constant $\beta_t = 1.5 \hspace{.1cm} \forall t$. 

\subsection{Bilevel Problem }
 Next, we compile the preceding sections into a single bilevel formulation. For compactness, variables and constraints are collected according to their corresponding level, so $ x_t := \left[ \alpha_t, \delta\alpha_t \right] \in \mathbb{R}^{2}\) and $ y_t := \left[ S_t, C_t, \Delta_t \right] \in \mathbb{R}^{3}$ are the upper-level (protocol) and lower-level (speculator) variables, respectively, at time $t$.

 Then, $\bm{x} := \left[ x_0, x_1, \hdots, x_T \right] \in \mathbb{R}^{2T} \mbox{ and } \bm{y} := \left[ y_0, y_1, \dots, y_T \right] \in \mathbb{R}^{3T}$ are the upper and lower variables concatenated over all steps $\timeT$.

The upper-level dynamics \cref{rates} are collected into the map $G : \mathbb{R}^{T} \rightarrow \mathbb{R}^{T}$
\begin{equation}
    \label{eqn:equalites}
    G(\bm{x},\bm{y}) := \begin{bmatrix}
        \alpha_{t+1} - \alpha_{t} - \delta\alpha_{t}
    \end{bmatrix}^{T}_{t=1} \in \mathbb{R}^{T}\,.
\end{equation}
Similarly, the lower-level dynamics \cref{supply}, \cref{collateral} map $g: \mathbb{R}^{2T} \rightarrow \mathbb{R}^{2T}$ is: 
\begin{equation}
\label{eqn:lower}
    g(\bm{x},\bm{y}) := \begin{bmatrix}
        \bigl[ S_{t} - S_{t-1} - \Delta_{t-1} \big]_{t=1}^{T}  \\
        \bigl[ C_{t} - C_{t-1} - \frac{\ps}{\pC}\Delta_{t-1}\bigr]_{t=1}^{T}
    \end{bmatrix} \in \mathbb{R}^{2T}
\end{equation}

and the vault constraint (\cref{vault}) 
\begin{equation}
    \label{eqn:equalites}
    h(\bm{x},\bm{y}) := \begin{bmatrix}
        \frac{C_t \pC}{S_t} - \beta_t
    \end{bmatrix}^{T}_{t=0} \in \mathbb{R}^{T}\,.
\end{equation}

Lastly, the upper and lower objectives are given by $ F(\bm{x},\bm{y}) := \JP  \mbox{ and } f(\bm{x},\bm{y}) := - \US $, where we negate $\US$
so that both agents are minimizing. We can now express the bilevel problem in compact notation.
\begin{subequations}
    \label{bilevel}
    \begin{alignat}{3}
          &(\bm{x^{*}},\bm{y^{*}}) \in &&\argmin_{\bm{x}, \tilde{\bm{y}}} && \hspace{.1 cm} F(\bm{x},\tilde{\bm{y}})  \\
          & && \qquad\textnormal{s.t.} && G(\bm{x},\tilde{\bm{y}}) = 0 \\ 
          \vspace{4mm}
          & && && \tilde{\bm{y}} \in \argmin_{\bm{y}} \hspace{.1cm} f(\bm{x},\bm{y}) \\ 
          & && && \qquad \qquad \textnormal{s.t.}~g(\bm{x},\bm{y}) = 0 \\
          & && && \qquad \qquad \hspace{0.5cm} h(\bm{x},\bm{y}) \geq 0 
    \end{alignat}
\end{subequations}
A solution $(\bm{x^{*}},\bm{y^{*} })$ to \cref{bilevel} is a \emph{Stackelberg equilibrium.} We note that, by setting $\ps = p^{\textnormal{Stb}}_0$ (the current market price observation), across all time steps in \cref{collateral}, the lower level problem in \cref{bilevel} becomes linear (and convex) in $y$. This slight approximation to problem \cref{bilevel} will ensure that the algorithm we employ for solving \cref{bilevel} finds a local minimizer \cite{bard2013practical,dempe2020bilevel}.


\subsection{Transforming Problem \cref{bilevel}}
It is well-known that bilevel formulations can be transformed into a single-level optimization problem by replacing the lower level problem in \cref{bilevel} with its Karush-Kuhn-Tucker (KKT) conditions \cite{bard2013practical, dempe2020bilevel}. To this end, we first introduce the speculator's \emph{Lagrangian}:
%
\begin{equation}
    \mathcal{L} = \JS - \blam^{\intercal}g(\bm{x},\bm{y}) - \bmu^{\intercal}h(\bm{x},\bm{y})
\end{equation}
where $\blam \in \mathbb{R}^{2T}$ and $\bmu \in \mathbb{R}^{T}$ are dual variables. Now, \cref{bilevel} can be rewritten as:
\begin{subequations}
\label{augmented}
\begin{alignat}{2}
    &\min_{\bm{x}, \bm{y}, \blam, \bmu}  && F(\bm{x}, \bm{y})  \\
    &\mbox{ subject to } \hspace{.2cm}  
          &&0 = G(\bm{x},\bm{y})  \\
    &     &&0 = \nabla_{\bm{y}} \mathcal{L}(\bm{x},\bm{y},\blam,\bmu)  \\
    &     &&0 = g(\bm{x},\bm{y})\\
    &     &&0 \leq \bmu \perp h(\bm{x},\bm{y}) \geq 0\,. \label{complement}
\end{alignat}
\end{subequations}

Line \cref{complement} makes Problem \cref{augmented} a special case of a broader class of problems known as \emph{mathematical programs with complementarity constraints} (MPCC) \cite{luo1996mathematical}. In practice, an MPCC can be difficult to solve due to the non-smoothness induced by the complementarity constraints. However, techniques for solving MPCC's are well-studied \cite{lin2005modified, guo2015solving, andreani2001solution}. In particular, one can introduce a slack variable $\beps \in \mathbb{R}^{T}$  \cite{guo2015solving} and replace \cref{augmented} with the following relaxation:
\begin{subequations}
\label{relaxed}
\begin{alignat}{2}
  &\min_{(\bm{x}, \bm{y},\blam, \bmu)} \hspace{.2cm } && F(\bm{x}, \bm{y})  \\
    &\mbox{ subject to } \hspace{.2cm } 
      && G(\bm{x},\bm{y}) = 0 ; \hspace{.1cm} g(\bm{x},\bm{y}) = 0  \\
    & && \nabla_{\bm{y}} \mathcal{L}(\bm{x},\bm{y},\blam,\bmu) = 0\\
    & && h(\bm{x},\bm{y}) \geq 0 ; \hspace{.1cm} \bm{\mu} \geq 0 \\
    & && \bmu^{\intercal}h(\bm{x},\bm{y}) - \beps = 0 \,.
\end{alignat}
\end{subequations}
We initialize $\beps = \begin{bmatrix} 1, & \dots &,1{}^{T-1} \end{bmatrix}$ and solve \cref{relaxed} iteratively, gradually reducing $\beps \rightarrow 0$ and tightening \cref{complement} at every step while using the solution from the prior iteration to warm-start the next. This will not always lead to an exact solution, but will often converge to an accurate approximation. The procedure is given explicitly within the inner loop of \cref{alg1}.

\begin{algorithm}
    \caption{Relaxed Receding Speculator Game} 
    \label{alg1}

    $S_{0}, C_{0}, \alpha_{0}, \widehat{p}^{S}_{0}, \widehat{p}^{C}_{0}, \tol , T \leftarrow$
        Initial State, Price Forecasts, Solver Tolerance, Time Horizon \\

    \While{$t < \infty$}{
      $\beps \leftarrow \bm{1}$  \tcp*{Reset $\epsilon$ } 
      \For{ \( n = 1 \) \textbf{ to } \( 10 \)}{
        $\bold{\zstar} = \begin{pmatrix} \bold{x}^{*}, \bold{y}^{*} \end{pmatrix} \leftarrow$ Solve relaxed Problem \cref{relaxed} \\
        \If{ $||\bold{\zstar} - \bold{\zlast}||_2 < \tol$}{
                $\alpha_t \leftarrow$  Select from first time step of \( \bold{x}^{*} \) 
                 \textbf{break}
            }
            \Else{
                $\beps \leftarrow \frac{1}{2}\beps$ \tcp*{Gradually reduce $\beps \rightarrow 0$}  
                $\bold{\zlast} \leftarrow \bold{\zstar}$
            }
      }
      $\Delta, p^{C}_{t}, p^{S}_{t} \leftarrow$ Get Market Observations \\
      $\widehat{p}^{C}_{t}, \widehat{p}^{S}_{t} \leftarrow $ Update forecasts from $p^{C}_{t}, p^{S}_{t}$ \\ 
      $S_{t}, C_{t} \leftarrow$ From $p^{C}_{t}, p^{S}_{t}, \Delta , \alpha$ using \cref{supply}, \cref{collateral}
    }
\end{algorithm}

\begin{figure*}[t]
\centerline{\includegraphics[width=\textwidth]{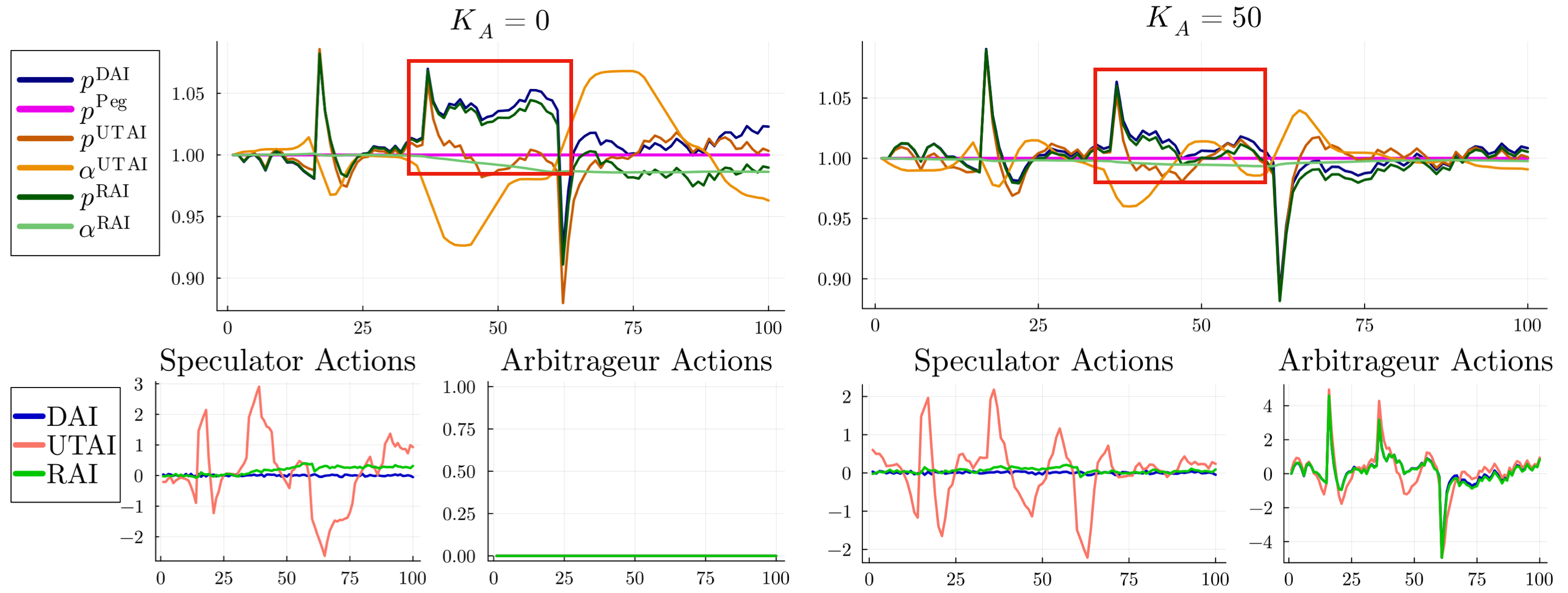}}
\caption{ \emph{Arbitrage Makes the Difference:} In the absence of arbitrage ($K_A=0$), RAI and DAI both struggle to regain their peg. When $K_A=50$ we see the arbitrageur actions correlate across all three tokens; however, only UTAI's redemption price decisions motivates unique speculator agent behavior which helps the system to regain the peg smoothly and within a shorter duration. In particular, note the trends highlighted within the red windows. UTAI regains the peg with or without arbitrage.}
\label{arbcompare}
\end{figure*}

\section{Experiments}

We benchmark the bilevel redemption price controller (which we label ``UTAI") against two baselines. The first of these (``DAI") fixes the redemption price at \$1. The second baseline (``RAI") adapts the redemption price according to equation \cref{pid}. 

\begin{figure}[b]
\centerline{\includegraphics[width=\columnwidth]{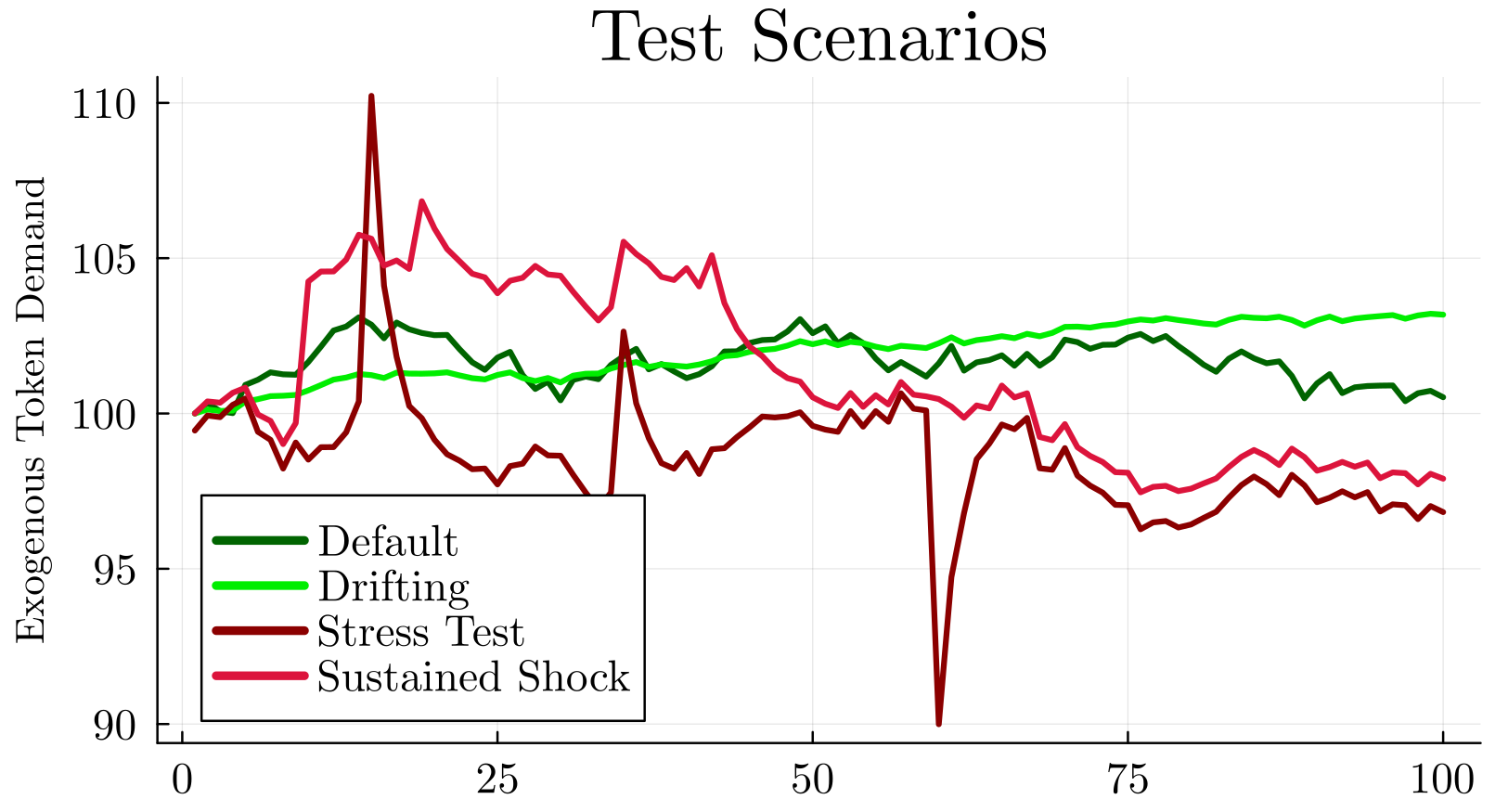}}
\caption{ A sample path from each scenario in \cref{section:scenarios}.}
\label{cases}
\end{figure}

\subsection{No Arbitrage}
We test in scenarios with both low and high volume of market arbitrage activity. This is because in a market with ample arbitrage and perfect efficiency, the \$1 constant redemption price is theoretically optimal, as arbitrageurs constantly eliminate price discrepancies. We expect UTAI to select a similar policy of relatively fixed redemption prices in this setting. Without arbitrageurs, on the other hand, the protocol will have to adapt redemption price more aggressively to incentivize speculators to keep the peg, through temporary changes to the cost of CDP maintenance. In low-arbitrage scenarios, DAI is especially vulnerable, and RAI will attempt to adapt, but we expect that neither can maintain a strong peg, especially in extreme conditions.


\subsection{Market Simulation}

The secondary market is modeled as consisting of these two primary forces: short-term arbitrageurs and longer-term CDP speculators who act as independent agents; in reality, these can be the same participants synchronously or asynchronously acting on different incentives. 

Agent actions are calculated as linear functions of their respective reference signals. Thus, the arbitrageur responds proportionally ($K_A$) to the difference between $\alpha$ (the redemption price) and $p$ (the market price), and the speculator responds proportionally ($K_S$) to both the change in redemption price as well as expected future ETH returns. Both agents use a time-disounted history of observations. Specifically:

\begin{align} \label{eqn:agents}
    \Delta^{A}_{t} &= K_{A}\sum_{\tau=0}^{\infty} \gamma^{t-\tau}\left(p^\mathrm{mkt}_{t}-\alpha_t \right)\\
    \Delta^{S}_{t} &= K_{S}\sum_{\tau=0}^{\infty} \gamma^{t-\tau}\left(p^\mathrm{mkt}_{t}-\alpha_t \right)
\end{align}

Agents add (or remove) $ \Delta^{A}_{t} + \Delta^{S}_{t}$ to the market supply at every step according to these rules. Market price is calculated as $\frac{D_t}{S_t}$ and the market clears at the target peg when $D_t=S_t$. Across experiments, $K_S$ is set to a constant value of 100 and  $K_A$ varies between $50$ or $0$ to ``activate" or ``deactivate" arbitrage. The values of $50$ and $100$ are chosen arbitrarily such that the maximum possible effects of the speculator and the arbitrageur in a single step are about equal.

\begin{figure*}[t]
\centerline{\includegraphics[width=\textwidth]{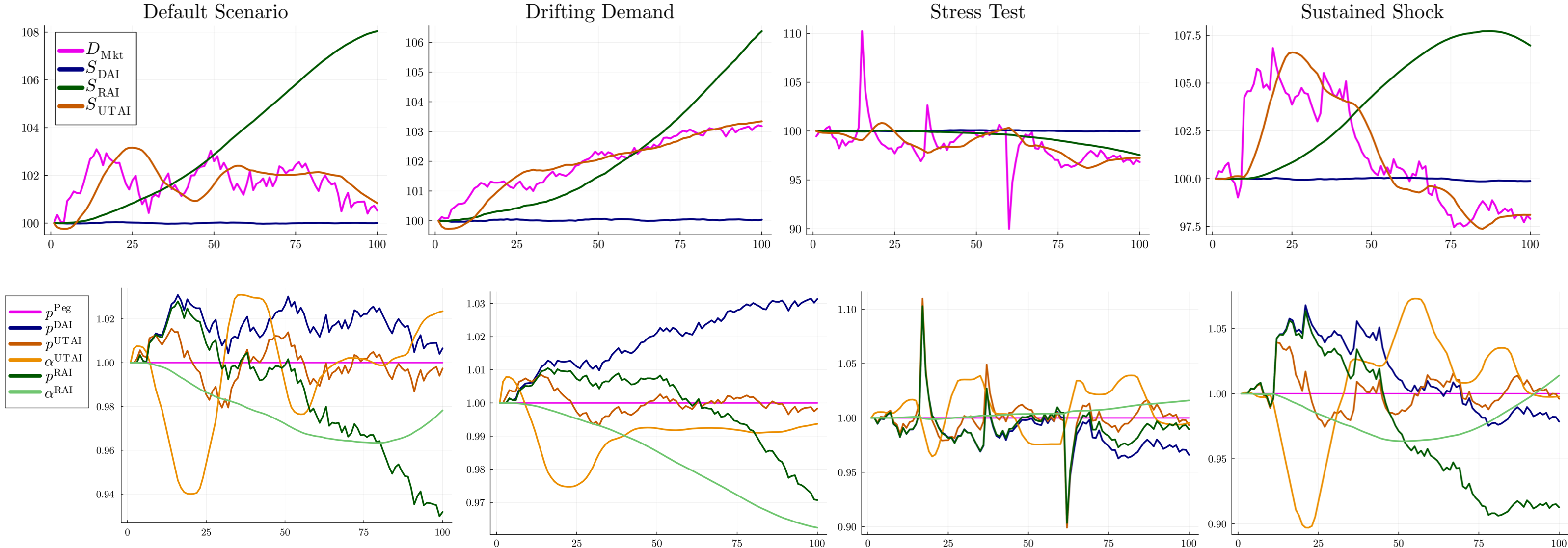}}
\caption{  Four sample runs exemplifying our four different scenarios selected from the ``no arbitrage" ($K_A = 0$) case. Each panel highlights unique behavior of UTAI under each condition. Specifically, the light orange series ($\alpha^{UTAI}$) are the controlled redemption prices which help guide the dark orange $p^{Stb}$ to the peg of \$1. Unlike RAI, which had to be re-tuned for the best results in each case/scenario, UTAI is naturally able to adapt to the scenario due to inherent reasoning through the predictive bilevel game model. In the absence of arbitrageurs, DAI's market price depegs and essentially tracks the exogenous demand.
}
\label{prices}
\end{figure*}

\subsection{Market Scenarios} \label{section:scenarios}

Each scenario is driven by a tuple of two exogenous random processes \((\bm{D}, \bm{p^\mathrm{Eth}})\) , representing token demand and Ethereum prices, both of which are simulated as geometric Brownian motions specified by volatility and drift parameters (\(\sigma\), \(\mu\)). The test cases are designed to cover a range of both extreme and benign conditions. \cref{cases} depicts a sample path from each case as a visual reference.

\subsubsection{Normal Operations (Default)}
The default case is characterized by relatively flat demand with some volatility in token demand. Minimal adjustments to the redemption price is expected to be optimal in this case, with the possible need for small corrections due to volatility.

\subsubsection{Drifting Demand}
This simulates a definitive linear trend of either increasing or decreasing exogenous token demand. As the bilevel protocol integrates forecasting information, it should anticipate the growth (or decline) and adjust the redemption price accordingly.

\subsubsection{Stress Test}
\label{stress test}
``Stress" is characterized by both a larger \(\sigma\) than the default case and several inserted point shocks  (2 positive and one negative) to the demand which decay exponentially. This simulates market `turbulence.' Shocks can be destabilizing, especially in low-arbitrage scenarios. We will be interested in seeing which token recovers the quickest and most gracefully.

\subsubsection{Sustained Shock} 
This case involves a sudden increase in demand which is sustained over a brief period of time, which then decays slowly, representing temporary one-sided price pressure. Again, this is expected to result in a depeg and we will be interested in observing recoveries.

\cref{prices} displays a single sample path from each of these 4 cases in a no-arbitrage environment, comparing the supply demand trends of each simulated token, as well the resulting market price trajectories. The results exemplify the nontrivial control decisions $\alpha^{UTAI}$ (in orange) versus $\alpha^{RAI}$ and the fixed redemption baseline.

The key takeaway from this is that UTAI can achieve the proper incentives to motivate speculator actions even in the absence of arbitrageurs. Introducing arbitrageurs back into the simulation only serves to improve all tokens, \emph{including} UTAI.
\cref{arbcompare} demonstrates the two scenarios side by side for \cref{stress test}. the effects of the arbitraging force on performance highlights the effects of the arbitraging force on performance in the stress test scenario.

\begin{figure*}[t]
\centerline{\includegraphics[width=0.9\textwidth,height=0.25\textwidth]{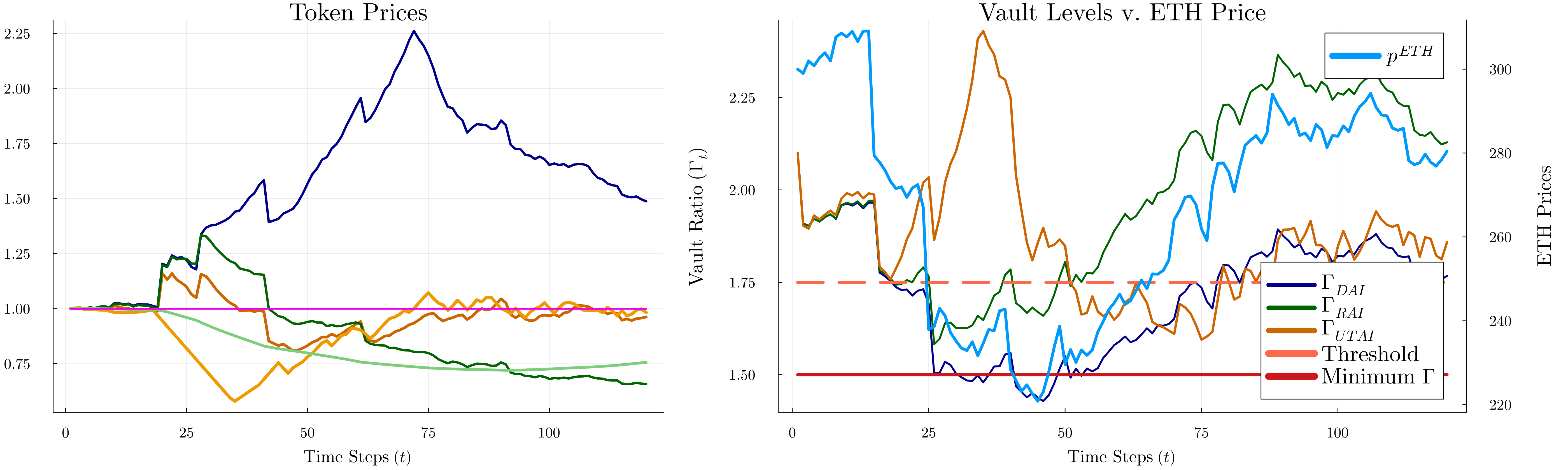}}
\caption{(Left) The demand shock triggered by token burning pressure leads to a deleveraging spiral in the price of DAI. RAI adapts but proportional control is slow to recover the peg. UTAI is quickest to recover. (Right) System collateral levels overlaid with a crashing ETH price pattern. The dashed red line depicts the threshold at which mass burning is activated in simulation; the bold red line shows the absolute minimum collateralization ratio. DAI performs worst, as any arbitraging is drowned out by the speculator's mass burn actions, and the system even falls below the minimum \(\beta\) requirement. Both RAI and UTAI mitigate deleveraging by adapting the redemption price, which effectively relaxes the vault constraint temporarily. UTAI stays furthest from the minimum collateralization threshold during the shock.}
\label{vaultplot}
\end{figure*}

\begin{table}[hb]
\huge
\centering
\captionsetup{justification=centering, font=small}
\caption{Monte Carlo Trials Summary}
\resizebox{\columnwidth}{!}{
\begin{tabular}{@{} ll ccc ccc ccc ccc @{}}  
    \toprule
    \textbf{Arbitrage Level} & \textbf{Scenario} 
    & N Samples 
    & \multicolumn{3}{c}{\emph{p-MAD}}
    & \multicolumn{2}{c}{\emph{r-MAD}} \\
    
    
     &   &  & \textbf{DAI} & \textbf{RAI} & \textbf{UTAI} & \textbf{RAI} & \textbf{UTAI} \\
    \midrule
    \multirow{4}{*}{$K_A = 0$}  
    & Default &  20 & 0.019417 & 0.022534 & \textbf{0.015386} & \textbf{0.038161} & 0.056736 \\
    & Demand Drift & 20 & 0.013664 & 0.006298 & \textbf{0.002304} & 0.012685 & \textbf{0.009242} \\
    & Stress Test & 20 & 0.027140 & 0.020917 & \textbf{0.012997} & 0.026070 & \textbf{0.033247}  \\
    & Sustained Shock & 20 & 0.030878 & 0.050162 & \textbf{0.012525} & 0.046472 & \textbf{0.045980}  \\
    \midrule
    \multirow{4}{*}{$K_A = 50$}  
    & Default     & 20 & 0.006178 & 0.006188 & \textbf{0.004522} & \textbf{0.006533} & 0.009078  \\
    & Demand Drift & 20 & 0.002804 & 0.002028 & \textbf{0.001682} & \textbf{0.002718} & 0.003005 \\
    & Stress Test  & 20 & 0.012895 & 0.013284 & \textbf{0.012052} & \textbf{0.013365} & 0.020706  \\
    & Sustained Shock & 20 & 0.049517 & 0.075299 & \textbf{0.024873} & 0.074763 & \textbf{0.068955}  \\
    \midrule
    &Avg. & 160 & 0.020051 & 0.023702 & \textbf{0.010421} & \textbf{0.026980} & 0.030126  \\ 
    &Median & 160 & 0.010998 & 0.008668 & \textbf{0.004630} & \textbf{0.012164} & 0.012259  \\ 
    &Std. Dev. & 160 &  0.023755 & 0.035049 & \textbf{0.018451} & \textbf{0.043770} & 0.058678 \\
    \bottomrule
\end{tabular}
}
\label{tbl:mc_stats}
\end{table}

\begin{figure}[h]
\centerline{\includegraphics[width=0.85\columnwidth]{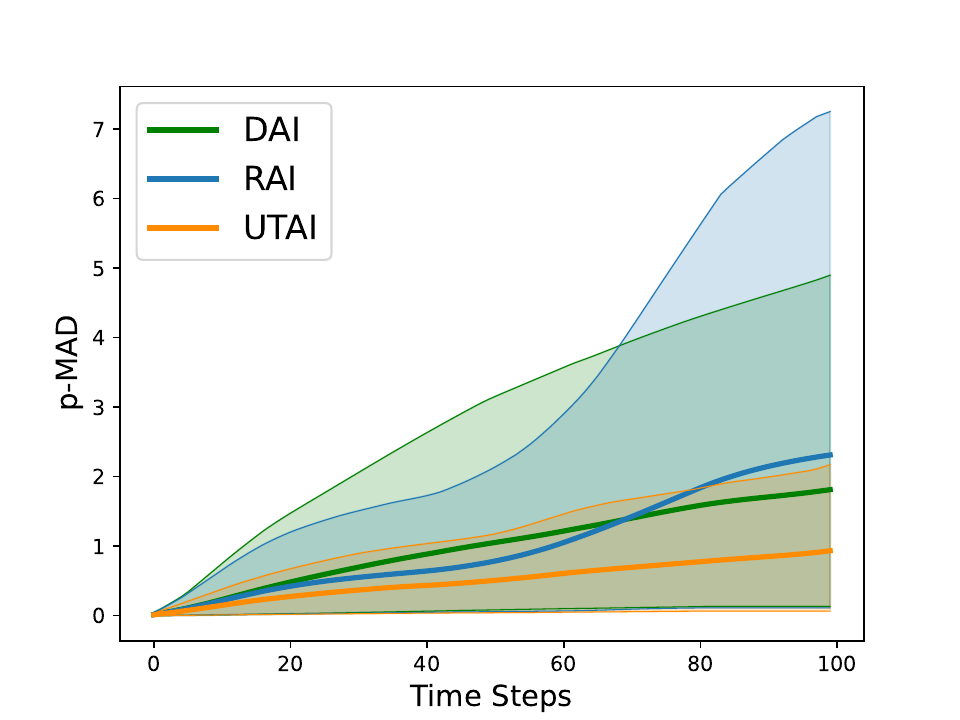}}
\caption{Results from \cref{tbl:mc_stats} pooled and plotted over the 100 time steps. Shading represents a 90\% confidence interval over the p-MAD metric.}
\label{mcplot}
\end{figure}

\subsection{Monte Carlo Study}

Next, we demonstrate the general performance of our method by conducting 20 trials over 100 time steps across the four scenarios both in the presence and absence of arbitrage. Performance is measured by mean absolute deviation from the \$1 peg (p-MAD) and mean absolute deviation of market price about the redemption price (r-MAD). The latter metric provides another way to compare to RAI since, as discussed in \cref{pid section}, Reflexer's stated goal is to align RAI with market prices and not necessarily peg to a specific value.

The results are summarized in \cref{tbl:mc_stats} and \cref{mcplot}. In short, UTAI achieves the best p-MAD in all cases, and closely approximates RAI's performance in the r-MAD metric. The way to interpret this is that, in our experiments, UTAI not only maintains the target peg but it does so without too much variation in its redemption price from that target peg. \cref{prices} depicts this fact where UTAI can be observed making flexible and timely redemption price changes before smoothly returning redemption back to par. RAI suffers in several categories due to the simplicity of proportional control, often leading to overshoot from which it can be slow to recover. 

This also explains why, in low-arbitrage scenarios, RAI may perform slightly better than DAI but ultimately still struggles. Our results indicate that, given a fixed $K_p$, RAI needs arbitraging trades nearly as much DAI and even with this, requires more sophisticated control if it is to target a specific peg. On the other hand, UTAI is highly flexible and resilient in every scenario regardless of arbitrage.

\subsection{Vault Safety Test}
\Cref{vaultplot} presents the results of our final experiment, which investigates how each token responds to a crisis event modeled after the ``deleveraging spiral'' discussed in \cref{deleveraging}. 
Specifically, this experiment tests whether UTAI can return the system to a sufficient collateralization level faster than the baselines while minimizing deviation from the price peg. 
In our market simulation, we include arbitrageurs (as discussed above), and amplify the speculator's burning behavior when collateral falls below a particular threshold (the red dashed line in the figure).
Observe how UTAI is immediately able to adapt to the perceived extreme conditions, and quickly steers the system away from the collateralization floor. This action mitigates deviation from the peg and facilitates eventual recovery.

\section{Conclusions \& Future Work}

In this paper, we have put forth a novel algorithmic redemption price controller for a decentralized crypto-backed stablecoin system based on dynamic game theory and optimal control.
Experiments show that our bilevel control design outperforms existing proportional control techniques and a fixed rate baseline, especially in extreme market conditions and in the absence of arbitrageurs.
Furthermore, our bilevel controller gives more precise autonomous pricing decisions to maintain a peg than the proportional controller baseline, which also must rely on offline re-tuning of the $K_p$ parameter \cite{summoning} as circumstances change. 
This is significant as it shows it can be possible for crypto-backed decentralized stablecoins while targeting a specific peg with the right controller.

\textbf{Discussion:} 
Although our results point to advantages of adopting more sophisticated control designs than those used in DeFi today, there are practical implementation details which still need to be considered. 
A key issue is a lack of sophisticated, faithful market simulations to more rigorously stress test new ideas for algorithmic DeFi mechanisms. 
To that end, we note that our model also provides utility as an off-line simulator for DeFi protocols, especially for system designers interested in algorithmic searches for setting optimal governance or in understanding microeconomic interactions and emergent behaviors. We believe there exists significant room in this domain to apply the rich tools of optimal control and dynamic game theory for both online and offline applications. 

\textbf{Future Work:} The framework we presented allows for several interesting extensions. One direction might be to consider a protocol with a more complex control space.
In this work, we modeled a single-collateral system, but it is now common for stablecoins to be backed by multiple cryptocurrencies.
For example, a recent work \cite{hajek2024collateral} used mean-variance optimization for selecting an optimal \emph{portfolio} of collateral assets, but highlights the need for a mechanism to achieve it in practice.
Our dynamic Stackelberg model naturally complements this approach and could be used to find the right incentives to build target collateral portfolios in reasonable time.

Another interesting direction is a data-driven variant of the problem. Here, we considered a \emph{forward} dynamic game model, in which the utility function of the participating agents is assumed to be known. One could instead investigate the \emph{inverse game} \cite{mehr2023maximum,li2023cost, peters2023online} in which the protocol attempts to \emph{learn} more precise agent utilities by filtering observations through a structured game model. 



\bibliographystyle{ACM-Reference-Format} 

\bibliography{refs}


\begin{thebibliography}{58}


\ifx \showCODEN    \undefined \def \showCODEN     #1{\unskip}     \fi
\ifx \showDOI      \undefined \def \showDOI       #1{#1}\fi
\ifx \showISBNx    \undefined \def \showISBNx     #1{\unskip}     \fi
\ifx \showISBNxiii \undefined \def \showISBNxiii  #1{\unskip}     \fi
\ifx \showISSN     \undefined \def \showISSN      #1{\unskip}     \fi
\ifx \showLCCN     \undefined \def \showLCCN      #1{\unskip}     \fi
\ifx \shownote     \undefined \def \shownote      #1{#1}          \fi
\ifx \showarticletitle \undefined \def \showarticletitle #1{#1}   \fi
\ifx \showURL      \undefined \def \showURL       {\relax}        \fi
\providecommand\bibfield[2]{#2}
\providecommand\bibinfo[2]{#2}
\providecommand\natexlab[1]{#1}
\providecommand\showeprint[2][]{arXiv:#2}

\bibitem[Adams and Ibert(2022)]%
        {adams2022runs}
\bibfield{author}{\bibinfo{person}{Austin Adams} {and} \bibinfo{person}{Markus Ibert}.} \bibinfo{year}{2022}\natexlab{}.
\newblock \showarticletitle{Runs on algorithmic stablecoins: Evidence from Iron, Titan, and Steel}.
\newblock  (\bibinfo{year}{2022}).
\newblock


\bibitem[Akcin et~al\mbox{.}(2022)]%
        {akcin2022control}
\bibfield{author}{\bibinfo{person}{Oguzhan Akcin}, \bibinfo{person}{Robert~P Streit}, \bibinfo{person}{Benjamin Oommen}, \bibinfo{person}{Sriram Vishwanath}, {and} \bibinfo{person}{Sandeep Chinchali}.} \bibinfo{year}{2022}\natexlab{}.
\newblock \showarticletitle{A control theoretic approach to infrastructure-centric blockchain tokenomics}.
\newblock \bibinfo{journal}{\emph{arXiv preprint arXiv:2210.12881}} (\bibinfo{year}{2022}).
\newblock


\bibitem[Allgower et~al\mbox{.}(2004)]%
        {allgower2004nonlinear}
\bibfield{author}{\bibinfo{person}{Frank Allgower}, \bibinfo{person}{Rolf Findeisen}, \bibinfo{person}{Zoltan~K Nagy}, {et~al\mbox{.}}} \bibinfo{year}{2004}\natexlab{}.
\newblock \showarticletitle{Nonlinear model predictive control: From theory to application}.
\newblock \bibinfo{journal}{\emph{Journal-Chinese Institute Of Chemical Engineers}} \bibinfo{volume}{35}, \bibinfo{number}{3} (\bibinfo{year}{2004}), \bibinfo{pages}{299--316}.
\newblock


\bibitem[Analysis(2023)]%
        {svb}
\bibfield{author}{\bibinfo{person}{Chain Analysis}.} \bibinfo{year}{2023}\natexlab{}.
\newblock \bibinfo{title}{Here’s What On-Chain Data Tells Us About Crypto’s Reaction to the Demise of Silicon Valley Bank And Its Impact on USDC}.
\newblock
\newblock
\urldef\tempurl%
\url{https://www.chainalysis.com/blog/crypto-market-usdc-silicon-valley-bank/}
\showURL{%
\tempurl}


\bibitem[Analytics(2023a)]%
        {moodya}
\bibfield{author}{\bibinfo{person}{Moody's Analytics}.} \bibinfo{year}{2023}\natexlab{a}.
\newblock \bibinfo{title}{Large fiat-backed stablecoins depegged 600+ times in 2023}.
\newblock
\newblock
\urldef\tempurl%
\url{https://www.moodys.com/web/en/us/insights/banking/moody-launches-new-digital-asset-monitor-to-track-risk.html}
\showURL{%
\tempurl}


\bibitem[Analytics(2023b)]%
        {moodyb}
\bibfield{author}{\bibinfo{person}{Moody's Analytics}.} \bibinfo{year}{2023}\natexlab{b}.
\newblock \bibinfo{title}{Stablecoins have been unstable. Why?}
\newblock
\newblock
\urldef\tempurl%
\url{https://www.moodys.com/web/en/us/about/insights/data-stories/stablecoins-instability.html}
\showURL{%
\tempurl}


\bibitem[Andreani and Mart{\i}{\'{}}~nez(2001)]%
        {andreani2001solution}
\bibfield{author}{\bibinfo{person}{Roberto Andreani} {and} \bibinfo{person}{Jos{\'e}~Mario Mart{\i}{\'{}}~nez}.} \bibinfo{year}{2001}\natexlab{}.
\newblock \showarticletitle{On the solution of mathematical programming problems with equilibrium constraints}.
\newblock \bibinfo{journal}{\emph{Mathematical Methods of Operations Research}}  \bibinfo{volume}{54} (\bibinfo{year}{2001}), \bibinfo{pages}{345--358}.
\newblock


\bibitem[Ante et~al\mbox{.}(2021)]%
        {ante2021influence}
\bibfield{author}{\bibinfo{person}{Lennart Ante}, \bibinfo{person}{Ingo Fiedler}, {and} \bibinfo{person}{Elias Strehle}.} \bibinfo{year}{2021}\natexlab{}.
\newblock \showarticletitle{The influence of stablecoin issuances on cryptocurrency markets}.
\newblock \bibinfo{journal}{\emph{Finance Research Letters}}  \bibinfo{volume}{41} (\bibinfo{year}{2021}), \bibinfo{pages}{101867}.
\newblock


\bibitem[Ariah Klages-Mundt(2021)]%
        {gyro}
\bibfield{author}{\bibinfo{person}{Daniel~Perez Ariah Klages-Mundt, Lewis~Gudgeon}.} \bibinfo{year}{2021}\natexlab{}.
\newblock \bibinfo{title}{Gryoscopic Stablecoins}.
\newblock
\newblock
\urldef\tempurl%
\url{https://github.com/gyrostable/gyroscope-landing/tree/master/pdfs}
\showURL{%
\tempurl}


\bibitem[Bard(2013)]%
        {bard2013practical}
\bibfield{author}{\bibinfo{person}{Jonathan~F Bard}.} \bibinfo{year}{2013}\natexlab{}.
\newblock \bibinfo{booktitle}{\emph{Practical bilevel optimization: algorithms and applications}}. Vol.~\bibinfo{volume}{30}.
\newblock \bibinfo{publisher}{Springer Science \& Business Media}.
\newblock


\bibitem[Bertucci et~al\mbox{.}(2024)]%
        {bertucci2024agents}
\bibfield{author}{\bibinfo{person}{Charles Bertucci}, \bibinfo{person}{Louis Bertucci}, \bibinfo{person}{Mathis Gontier~Delaunay}, \bibinfo{person}{Olivier Gueant}, {and} \bibinfo{person}{Matthieu Lesbre}.} \bibinfo{year}{2024}\natexlab{}.
\newblock \showarticletitle{Agents' Behavior and Interest Rate Model Optimization in DeFi Lending}.
\newblock \bibinfo{journal}{\emph{Available at SSRN 4802776}} (\bibinfo{year}{2024}).
\newblock


\bibitem[BlockScience(2021)]%
        {summoning}
\bibfield{author}{\bibinfo{person}{BlockScience}.} \bibinfo{year}{2021}\natexlab{}.
\newblock \bibinfo{title}{Summoning the Money God}.
\newblock
\newblock
\urldef\tempurl%
\url{https://medium.com/reflexer-labs/summoning-the-money-god-2a3f3564a5f2}
\showURL{%
\tempurl}


\bibitem[Brotcorne et~al\mbox{.}(2001)]%
        {brotcorne2001bilevel}
\bibfield{author}{\bibinfo{person}{Luce Brotcorne}, \bibinfo{person}{Martine Labb{\'e}}, \bibinfo{person}{Patrice Marcotte}, {and} \bibinfo{person}{Gilles Savard}.} \bibinfo{year}{2001}\natexlab{}.
\newblock \showarticletitle{A bilevel model for toll optimization on a multicommodity transportation network}.
\newblock \bibinfo{journal}{\emph{Transportation science}} \bibinfo{volume}{35}, \bibinfo{number}{4} (\bibinfo{year}{2001}), \bibinfo{pages}{345--358}.
\newblock


\bibitem[Chitra et~al\mbox{.}(2022)]%
        {chitra2022defi}
\bibfield{author}{\bibinfo{person}{Tarun Chitra}, \bibinfo{person}{Kshitij Kulkarni}, \bibinfo{person}{Guillermo Angeris}, \bibinfo{person}{Alex Evans}, {and} \bibinfo{person}{Victor Xu}.} \bibinfo{year}{2022}\natexlab{}.
\newblock \bibinfo{title}{Defi liquidity management via optimal control: ohm as a case study}.
\newblock
\newblock


\bibitem[Clements(2021)]%
        {clements2021built}
\bibfield{author}{\bibinfo{person}{Ryan Clements}.} \bibinfo{year}{2021}\natexlab{}.
\newblock \showarticletitle{Built to fail: The inherent fragility of algorithmic stablecoins}.
\newblock \bibinfo{journal}{\emph{Wake Forest L. Rev. Online}}  \bibinfo{volume}{11} (\bibinfo{year}{2021}), \bibinfo{pages}{131}.
\newblock


\bibitem[CoinMarketCap(2024)]%
        {coinmarket}
\bibfield{author}{\bibinfo{person}{CoinMarketCap}.} \bibinfo{year}{2024}\natexlab{}.
\newblock \bibinfo{title}{Top Stablecoin Tokens by Market Capitalization}.
\newblock
\newblock
\urldef\tempurl%
\url{https://coinmarketcap.com/view/stablecoin/}
\showURL{%
\tempurl}


\bibitem[Dao(2021)]%
        {psm}
\bibfield{author}{\bibinfo{person}{Maker Dao}.} \bibinfo{year}{2021}\natexlab{}.
\newblock \bibinfo{title}{MIP29: Peg Stability Module}.
\newblock
\newblock
\urldef\tempurl%
\url{https://mips.makerdao.com/mips/details/MIP29}
\showURL{%
\tempurl}


\bibitem[Dempe and Zemkoho(2020)]%
        {dempe2020bilevel}
\bibfield{author}{\bibinfo{person}{Stephan Dempe} {and} \bibinfo{person}{Alain Zemkoho}.} \bibinfo{year}{2020}\natexlab{}.
\newblock \showarticletitle{Bilevel optimization}.
\newblock In \bibinfo{booktitle}{\emph{Springer optimization and its applications}}. Vol.~\bibinfo{volume}{161}. \bibinfo{publisher}{Springer}.
\newblock


\bibitem[d’Avernas et~al\mbox{.}(2021)]%
        {d2021stablecoins}
\bibfield{author}{\bibinfo{person}{Adrien d’Avernas}, \bibinfo{person}{Thomas Bourany}, {and} \bibinfo{person}{Quentin Vandeweyer}.} \bibinfo{year}{2021}\natexlab{}.
\newblock \bibinfo{title}{Are stablecoins stable?}
\newblock
\newblock
\urldef\tempurl%
\url{https://www. banque-france. fr/sites/default/files/media/2021/06/10/gdre\_ bounary. pdf}
\showURL{%
\tempurl}


\bibitem[Finance(2023)]%
        {euler}
\bibfield{author}{\bibinfo{person}{Euler Finance}.} \bibinfo{year}{2023}\natexlab{}.
\newblock \bibinfo{title}{White Paper}.
\newblock
\newblock
\urldef\tempurl%
\url{https://docs-v1.euler.finance/getting-started/white-paper}
\showURL{%
\tempurl}


\bibitem[Garcia et~al\mbox{.}(1989)]%
        {garcia1989model}
\bibfield{author}{\bibinfo{person}{Carlos~E Garcia}, \bibinfo{person}{David~M Prett}, {and} \bibinfo{person}{Manfred Morari}.} \bibinfo{year}{1989}\natexlab{}.
\newblock \showarticletitle{Model predictive control: Theory and practice—A survey}.
\newblock \bibinfo{journal}{\emph{Automatica}} \bibinfo{volume}{25}, \bibinfo{number}{3} (\bibinfo{year}{1989}), \bibinfo{pages}{335--348}.
\newblock


\bibitem[Gauntlet(2020)]%
        {feedback}
\bibfield{author}{\bibinfo{person}{Gauntlet}.} \bibinfo{year}{2020}\natexlab{}.
\newblock \bibinfo{title}{Feedback Control as a New Primitive}.
\newblock
\newblock
\urldef\tempurl%
\url{https://medium.com/gauntlet-networks/feedback-control-as-a-new-primitive-for-defi-27b493f25b1}
\showURL{%
\tempurl}


\bibitem[Grobys et~al\mbox{.}(2021)]%
        {grobys2021stability}
\bibfield{author}{\bibinfo{person}{Klaus Grobys}, \bibinfo{person}{Juha Junttila}, \bibinfo{person}{James~W Kolari}, {and} \bibinfo{person}{Niranjan Sapkota}.} \bibinfo{year}{2021}\natexlab{}.
\newblock \showarticletitle{On the stability of stablecoins}.
\newblock \bibinfo{journal}{\emph{Journal of Empirical Finance}}  \bibinfo{volume}{64} (\bibinfo{year}{2021}), \bibinfo{pages}{207--223}.
\newblock


\bibitem[Guo et~al\mbox{.}(2015)]%
        {guo2015solving}
\bibfield{author}{\bibinfo{person}{Lei Guo}, \bibinfo{person}{Gui-Hua Lin}, {and} \bibinfo{person}{Jane~J Ye}.} \bibinfo{year}{2015}\natexlab{}.
\newblock \showarticletitle{Solving mathematical programs with equilibrium constraints}.
\newblock \bibinfo{journal}{\emph{Journal of Optimization Theory and Applications}}  \bibinfo{volume}{166} (\bibinfo{year}{2015}), \bibinfo{pages}{234--256}.
\newblock


\bibitem[Hajek et~al\mbox{.}(2024)]%
        {hajek2024collateral}
\bibfield{author}{\bibinfo{person}{Bretislav Hajek}, \bibinfo{person}{Daniel Reijsbergen}, \bibinfo{person}{Anwitaman Datta}, {and} \bibinfo{person}{Jussi Keppo}.} \bibinfo{year}{2024}\natexlab{}.
\newblock \showarticletitle{Collateral Portfolio Optimization in Crypto-Backed Stablecoins}.
\newblock \bibinfo{journal}{\emph{arXiv preprint arXiv:2405.08305}} (\bibinfo{year}{2024}).
\newblock


\bibitem[Herzog et~al\mbox{.}(2007)]%
        {herzog2007stochastic}
\bibfield{author}{\bibinfo{person}{Florian Herzog}, \bibinfo{person}{Gabriel Dondi}, {and} \bibinfo{person}{Hans~P Geering}.} \bibinfo{year}{2007}\natexlab{}.
\newblock \showarticletitle{Stochastic model predictive control and portfolio optimization}.
\newblock \bibinfo{journal}{\emph{International Journal of Theoretical and Applied Finance}} \bibinfo{volume}{10}, \bibinfo{number}{02} (\bibinfo{year}{2007}), \bibinfo{pages}{203--233}.
\newblock


\bibitem[Herzog et~al\mbox{.}(2006)]%
        {herzog2006model}
\bibfield{author}{\bibinfo{person}{Florian Herzog}, \bibinfo{person}{Simon Keel}, \bibinfo{person}{Gabriel Dondi}, \bibinfo{person}{Lorenz~M Schumann}, {and} \bibinfo{person}{Hans~P Geering}.} \bibinfo{year}{2006}\natexlab{}.
\newblock \showarticletitle{Model predictive control for portfolio selection}. In \bibinfo{booktitle}{\emph{2006 American Control Conference}}. IEEE, \bibinfo{pages}{8--pp}.
\newblock


\bibitem[Jensen et~al\mbox{.}(2021)]%
        {jensen2021introduction}
\bibfield{author}{\bibinfo{person}{Johannes~Rude Jensen}, \bibinfo{person}{Victor von Wachter}, {and} \bibinfo{person}{Omri Ross}.} \bibinfo{year}{2021}\natexlab{}.
\newblock \showarticletitle{An introduction to decentralized finance (defi)}.
\newblock \bibinfo{journal}{\emph{Complex Systems Informatics and Modeling Quarterly}} \bibinfo{number}{26} (\bibinfo{year}{2021}), \bibinfo{pages}{46--54}.
\newblock


\bibitem[Josefsson and Patriksson(2007)]%
        {josefsson2007sensitivity}
\bibfield{author}{\bibinfo{person}{Magnus Josefsson} {and} \bibinfo{person}{Michael Patriksson}.} \bibinfo{year}{2007}\natexlab{}.
\newblock \showarticletitle{Sensitivity analysis of separable traffic equilibrium equilibria with application to bilevel optimization in network design}.
\newblock \bibinfo{journal}{\emph{Transportation Research Part B: Methodological}} \bibinfo{volume}{41}, \bibinfo{number}{1} (\bibinfo{year}{2007}), \bibinfo{pages}{4--31}.
\newblock


\bibitem[Kalashnikov et~al\mbox{.}(2015)]%
        {kalashnikov2015bilevel}
\bibfield{author}{\bibinfo{person}{Vyacheslav~V Kalashnikov}, \bibinfo{person}{Stephan Dempe}, \bibinfo{person}{Gerardo~A P{\'e}rez-Vald{\'e}s}, \bibinfo{person}{Nataliya~I Kalashnykova}, \bibinfo{person}{Jos{\'e}-Fernando Camacho-Vallejo}, {et~al\mbox{.}}} \bibinfo{year}{2015}\natexlab{}.
\newblock \showarticletitle{Bilevel programming and applications}.
\newblock \bibinfo{journal}{\emph{Mathematical Problems in Engineering}}  \bibinfo{volume}{2015} (\bibinfo{year}{2015}).
\newblock


\bibitem[Kazemian et~al\mbox{.}(2022)]%
        {kazemian2022frax}
\bibfield{author}{\bibinfo{person}{Sam Kazemian}, \bibinfo{person}{Jason Huan}, \bibinfo{person}{Jonathan Shomroni}, {and} \bibinfo{person}{Kedar Iyer}.} \bibinfo{year}{2022}\natexlab{}.
\newblock \showarticletitle{Frax: A Fractional-Algorithmic Stablecoin Protocol}. In \bibinfo{booktitle}{\emph{2022 IEEE International Conference on Blockchain (Blockchain)}}. IEEE, \bibinfo{pages}{406--411}.
\newblock


\bibitem[Klages-Mundt et~al\mbox{.}(2020)]%
        {klages2020stablecoins}
\bibfield{author}{\bibinfo{person}{Ariah Klages-Mundt}, \bibinfo{person}{Dominik Harz}, \bibinfo{person}{Lewis Gudgeon}, \bibinfo{person}{Jun-You Liu}, {and} \bibinfo{person}{Andreea Minca}.} \bibinfo{year}{2020}\natexlab{}.
\newblock \showarticletitle{Stablecoins 2.0: Economic foundations and risk-based models}. In \bibinfo{booktitle}{\emph{Proceedings of the 2nd ACM Conference on Advances in Financial Technologies}}. \bibinfo{pages}{59--79}.
\newblock


\bibitem[Klages-Mundt and Minca(2021)]%
        {klages2021stability}
\bibfield{author}{\bibinfo{person}{Ariah Klages-Mundt} {and} \bibinfo{person}{Andreea Minca}.} \bibinfo{year}{2021}\natexlab{}.
\newblock \showarticletitle{(In)stability for the blockchain: Deleveraging spirals and stablecoin attacks}.
\newblock  (\bibinfo{year}{2021}).
\newblock


\bibitem[Klages-Mundt and Minca(2022)]%
        {klages2022while}
\bibfield{author}{\bibinfo{person}{Ariah Klages-Mundt} {and} \bibinfo{person}{Andreea Minca}.} \bibinfo{year}{2022}\natexlab{}.
\newblock \showarticletitle{While stability lasts: A stochastic model of noncustodial stablecoins}.
\newblock \bibinfo{journal}{\emph{Mathematical Finance}} \bibinfo{volume}{32}, \bibinfo{number}{4} (\bibinfo{year}{2022}), \bibinfo{pages}{943--981}.
\newblock


\bibitem[Labb{\'e} and Violin(2016)]%
        {labbe2016bilevel}
\bibfield{author}{\bibinfo{person}{Martine Labb{\'e}} {and} \bibinfo{person}{Alessia Violin}.} \bibinfo{year}{2016}\natexlab{}.
\newblock \showarticletitle{Bilevel programming and price setting problems}.
\newblock \bibinfo{journal}{\emph{Annals of operations research}}  \bibinfo{volume}{240} (\bibinfo{year}{2016}), \bibinfo{pages}{141--169}.
\newblock


\bibitem[Labs({[n.\,d.]})]%
        {rai}
\bibfield{author}{\bibinfo{person}{Reflexer Labs}.} \bibinfo{year}{[n.\,d.]}\natexlab{}.
\newblock \bibinfo{title}{RAI Finance}.
\newblock
\newblock
\urldef\tempurl%
\url{https://rai.finance/RAI-Finance-WhitePaper.pdf}
\showURL{%
\tempurl}


\bibitem[Lauko and Pardoe(2021)]%
        {lauko2021liquity}
\bibfield{author}{\bibinfo{person}{Robert Lauko} {and} \bibinfo{person}{Richard Pardoe}.} \bibinfo{year}{2021}\natexlab{}.
\newblock \bibinfo{booktitle}{\emph{Liquity: Decentralized borrowing protocol}}.
\newblock \bibinfo{type}{{T}echnical {R}eport}. \bibinfo{institution}{Technical report. Available at https://docsend. com/view/bwiczmy}.
\newblock


\bibitem[Li et~al\mbox{.}(2023)]%
        {li2023cost}
\bibfield{author}{\bibinfo{person}{Jingqi Li}, \bibinfo{person}{Chih-Yuan Chiu}, \bibinfo{person}{Lasse Peters}, \bibinfo{person}{Somayeh Sojoudi}, \bibinfo{person}{Claire Tomlin}, {and} \bibinfo{person}{David Fridovich-Keil}.} \bibinfo{year}{2023}\natexlab{}.
\newblock \showarticletitle{Cost inference for feedback dynamic games from noisy partial state observations and incomplete trajectories}.
\newblock \bibinfo{journal}{\emph{arXiv preprint arXiv:2301.01398}} (\bibinfo{year}{2023}).
\newblock


\bibitem[Li and Sethi(2017)]%
        {li2017review}
\bibfield{author}{\bibinfo{person}{Tao Li} {and} \bibinfo{person}{Suresh~P Sethi}.} \bibinfo{year}{2017}\natexlab{}.
\newblock \showarticletitle{A review of dynamic Stackelberg game models}.
\newblock \bibinfo{journal}{\emph{Discrete \& Continuous Dynamical Systems-B}} \bibinfo{volume}{22}, \bibinfo{number}{1} (\bibinfo{year}{2017}), \bibinfo{pages}{125}.
\newblock


\bibitem[Lin and Fukushima(2005)]%
        {lin2005modified}
\bibfield{author}{\bibinfo{person}{Gui-Hua Lin} {and} \bibinfo{person}{Masao Fukushima}.} \bibinfo{year}{2005}\natexlab{}.
\newblock \showarticletitle{A modified relaxation scheme for mathematical programs with complementarity constraints}.
\newblock \bibinfo{journal}{\emph{Annals of Operations Research}} \bibinfo{volume}{133}, \bibinfo{number}{1} (\bibinfo{year}{2005}), \bibinfo{pages}{63--84}.
\newblock


\bibitem[Liu et~al\mbox{.}(2023)]%
        {liu2023anatomy}
\bibfield{author}{\bibinfo{person}{Jiageng Liu}, \bibinfo{person}{Igor Makarov}, {and} \bibinfo{person}{Antoinette Schoar}.} \bibinfo{year}{2023}\natexlab{}.
\newblock \bibinfo{booktitle}{\emph{Anatomy of a run: The terra luna crash}}.
\newblock \bibinfo{type}{{T}echnical {R}eport}. \bibinfo{institution}{National Bureau of Economic Research}.
\newblock


\bibitem[Luo et~al\mbox{.}(1996)]%
        {luo1996mathematical}
\bibfield{author}{\bibinfo{person}{Zhi-Quan Luo}, \bibinfo{person}{Jong-Shi Pang}, {and} \bibinfo{person}{Daniel Ralph}.} \bibinfo{year}{1996}\natexlab{}.
\newblock \bibinfo{booktitle}{\emph{Mathematical programs with equilibrium constraints}}.
\newblock \bibinfo{publisher}{Cambridge University Press}.
\newblock


\bibitem[Marcotte(1986)]%
        {marcotte1986network}
\bibfield{author}{\bibinfo{person}{Patrice Marcotte}.} \bibinfo{year}{1986}\natexlab{}.
\newblock \showarticletitle{Network design problem with congestion effects: A case of bilevel programming}.
\newblock \bibinfo{journal}{\emph{Mathematical programming}} \bibinfo{volume}{34}, \bibinfo{number}{2} (\bibinfo{year}{1986}), \bibinfo{pages}{142--162}.
\newblock


\bibitem[MARS(2021)]%
        {mars}
\bibfield{author}{\bibinfo{person}{MARS}.} \bibinfo{year}{2021}\natexlab{}.
\newblock \bibinfo{title}{LitePaper}.
\newblock
\newblock
\urldef\tempurl%
\url{https://blog.marsprotocol.io/blog/mars-protocol-litepaper-2-0}
\showURL{%
\tempurl}


\bibitem[Mehr et~al\mbox{.}(2023)]%
        {mehr2023maximum}
\bibfield{author}{\bibinfo{person}{Negar Mehr}, \bibinfo{person}{Mingyu Wang}, \bibinfo{person}{Maulik Bhatt}, {and} \bibinfo{person}{Mac Schwager}.} \bibinfo{year}{2023}\natexlab{}.
\newblock \showarticletitle{Maximum-entropy multi-agent dynamic games: Forward and inverse solutions}.
\newblock \bibinfo{journal}{\emph{IEEE transactions on robotics}} \bibinfo{volume}{39}, \bibinfo{number}{3} (\bibinfo{year}{2023}), \bibinfo{pages}{1801--1815}.
\newblock


\bibitem[Mita et~al\mbox{.}(2019)]%
        {mita2019stablecoin}
\bibfield{author}{\bibinfo{person}{Makiko Mita}, \bibinfo{person}{Kensuke Ito}, \bibinfo{person}{Shohei Ohsawa}, {and} \bibinfo{person}{Hideyuki Tanaka}.} \bibinfo{year}{2019}\natexlab{}.
\newblock \showarticletitle{What is stablecoin?: A survey on price stabilization mechanisms for decentralized payment systems}. In \bibinfo{booktitle}{\emph{2019 8th International Congress on Advanced Applied Informatics (IIAI-AAI)}}. IEEE, \bibinfo{pages}{60--66}.
\newblock


\bibitem[Peters et~al\mbox{.}(2023)]%
        {peters2023online}
\bibfield{author}{\bibinfo{person}{Lasse Peters}, \bibinfo{person}{Vicen{\c{c}} Rubies-Royo}, \bibinfo{person}{Claire~J Tomlin}, \bibinfo{person}{Laura Ferranti}, \bibinfo{person}{Javier Alonso-Mora}, \bibinfo{person}{Cyrill Stachniss}, {and} \bibinfo{person}{David Fridovich-Keil}.} \bibinfo{year}{2023}\natexlab{}.
\newblock \showarticletitle{Online and offline learning of player objectives from partial observations in dynamic games}.
\newblock \bibinfo{journal}{\emph{The International Journal of Robotics Research}} \bibinfo{volume}{42}, \bibinfo{number}{10} (\bibinfo{year}{2023}), \bibinfo{pages}{917--937}.
\newblock


\bibitem[Qin and Badgwell(2003)]%
        {qin2003survey}
\bibfield{author}{\bibinfo{person}{S~Joe Qin} {and} \bibinfo{person}{Thomas~A Badgwell}.} \bibinfo{year}{2003}\natexlab{}.
\newblock \showarticletitle{A survey of industrial model predictive control technology}.
\newblock \bibinfo{journal}{\emph{Control engineering practice}} \bibinfo{volume}{11}, \bibinfo{number}{7} (\bibinfo{year}{2003}), \bibinfo{pages}{733--764}.
\newblock


\bibitem[S{\'a}nchez(2019)]%
        {sanchez2019truthful}
\bibfield{author}{\bibinfo{person}{David~Cerezo S{\'a}nchez}.} \bibinfo{year}{2019}\natexlab{}.
\newblock \showarticletitle{Truthful and faithful monetary policy for a stablecoin conducted by a decentralised, encrypted artificial intelligence}.
\newblock \bibinfo{journal}{\emph{arXiv preprint arXiv:1909.07445}} (\bibinfo{year}{2019}).
\newblock


\bibitem[Schueffel(2021)]%
        {schueffel2021defi}
\bibfield{author}{\bibinfo{person}{Patrick Schueffel}.} \bibinfo{year}{2021}\natexlab{}.
\newblock \showarticletitle{Defi: Decentralized finance-an introduction and overview}.
\newblock \bibinfo{journal}{\emph{Journal of Innovation Management}} \bibinfo{volume}{9}, \bibinfo{number}{3} (\bibinfo{year}{2021}), \bibinfo{pages}{I--XI}.
\newblock


\bibitem[Schwenzer et~al\mbox{.}(2021)]%
        {schwenzer2021review}
\bibfield{author}{\bibinfo{person}{Max Schwenzer}, \bibinfo{person}{Muzaffer Ay}, \bibinfo{person}{Thomas Bergs}, {and} \bibinfo{person}{Dirk Abel}.} \bibinfo{year}{2021}\natexlab{}.
\newblock \showarticletitle{Review on model predictive control: An engineering perspective}.
\newblock \bibinfo{journal}{\emph{The International Journal of Advanced Manufacturing Technology}} \bibinfo{volume}{117}, \bibinfo{number}{5} (\bibinfo{year}{2021}), \bibinfo{pages}{1327--1349}.
\newblock


\bibitem[Team(2017)]%
        {team2017dai}
\bibfield{author}{\bibinfo{person}{Maker Team}.} \bibinfo{year}{2017}\natexlab{}.
\newblock \bibinfo{title}{The Dai Stablecoin System Whitepaper}.
\newblock
\newblock
\urldef\tempurl%
\url{https://makerdao.com/whitepaper/DaiDec17WP.pdf}
\showURL{%
\tempurl}


\bibitem[Team(2018)]%
        {nubits}
\bibfield{author}{\bibinfo{person}{Reserve~Research Team}.} \bibinfo{year}{2018}\natexlab{}.
\newblock \bibinfo{title}{The End of a Stablecoin — The Case of NuBits}.
\newblock
\newblock
\urldef\tempurl%
\url{https://blog.reserve.org/the-end-of-a-stablecoin-the-case-of-nubits-dd1f0fb427a9}
\showURL{%
\tempurl}


\bibitem[Trimborn et~al\mbox{.}(2017)]%
        {trimborn2017portfolio}
\bibfield{author}{\bibinfo{person}{Torsten Trimborn}, \bibinfo{person}{Lorenzo Pareschi}, {and} \bibinfo{person}{Martin Frank}.} \bibinfo{year}{2017}\natexlab{}.
\newblock \showarticletitle{Portfolio optimization and model predictive control: a kinetic approach}.
\newblock \bibinfo{journal}{\emph{arXiv preprint arXiv:1711.03291}} (\bibinfo{year}{2017}).
\newblock


\bibitem[Von~Stackelberg(2010)]%
        {von2010market}
\bibfield{author}{\bibinfo{person}{Heinrich Von~Stackelberg}.} \bibinfo{year}{2010}\natexlab{}.
\newblock \bibinfo{booktitle}{\emph{Market structure and equilibrium}}.
\newblock \bibinfo{publisher}{Springer Science \& Business Media}.
\newblock


\bibitem[Werner et~al\mbox{.}(2022)]%
        {werner2022sok}
\bibfield{author}{\bibinfo{person}{Sam Werner}, \bibinfo{person}{Daniel Perez}, \bibinfo{person}{Lewis Gudgeon}, \bibinfo{person}{Ariah Klages-Mundt}, \bibinfo{person}{Dominik Harz}, {and} \bibinfo{person}{William Knottenbelt}.} \bibinfo{year}{2022}\natexlab{}.
\newblock \showarticletitle{Sok: Decentralized finance (defi)}. In \bibinfo{booktitle}{\emph{Proceedings of the 4th ACM Conference on Advances in Financial Technologies}}. \bibinfo{pages}{30--46}.
\newblock


\bibitem[Yu and Hong(2016)]%
        {yu2016supply}
\bibfield{author}{\bibinfo{person}{Mengmeng Yu} {and} \bibinfo{person}{Seung~Ho Hong}.} \bibinfo{year}{2016}\natexlab{}.
\newblock \showarticletitle{Supply--demand balancing for power management in smart grid: A Stackelberg game approach}.
\newblock \bibinfo{journal}{\emph{Applied energy}}  \bibinfo{volume}{164} (\bibinfo{year}{2016}), \bibinfo{pages}{702--710}.
\newblock


\bibitem[Zugno et~al\mbox{.}(2013)]%
        {zugno2013bilevel}
\bibfield{author}{\bibinfo{person}{Marco Zugno}, \bibinfo{person}{Juan~Miguel Morales}, \bibinfo{person}{Pierre Pinson}, {and} \bibinfo{person}{Henrik Madsen}.} \bibinfo{year}{2013}\natexlab{}.
\newblock \showarticletitle{A bilevel model for electricity retailers' participation in a demand response market environment}.
\newblock \bibinfo{journal}{\emph{Energy Economics}}  \bibinfo{volume}{36} (\bibinfo{year}{2013}), \bibinfo{pages}{182--197}.
\newblock


\end{thebibliography}


\end{document}